\documentstyle[preprint,aps,floats]{revtex}
\tightenlines

\input{epsf.tex}
\begin{document}
%\rightline{DAMTP-1999-50}
\def\ba{\begin{eqnarray}}
\def\ea{\end{eqnarray}}
\def\be{\begin{equation}}
\def\ee{\end{equation}}
\def\tr{{\rm tr}}

\title{A Braneworld Universe From Colliding Bubbles}

\author{Martin Bucher\thanks{E-mail: M.A.Bucher@damtp.cam.ac.uk}\\
DAMTP, Centre for Mathematical Sciences, University of Cambridge\\
Wilberforce Road, Cambridge CB3 0WA, United Kingdom}

\date{18 July 2001}

\maketitle

\begin{abstract}%
Much work has been devoted to the phenomenology 
and cosmology of the  so-called braneworld universe, 
where the $(3+1)$-dimensional universe familiar to us lies on 
a brane surrounded by a (4+1)-dimensional bulk spacetime
that is essentially empty except for a negative cosmological 
constant and the various modes associated with gravity.
For such a braneworld cosmology, the difficulty of
justifying a set of preferred initial conditions 
inevitably arises. The various proposals for 
inflation restricted to the brane only partially explain 
the homogeneity and isotropy of the resulting braneworld universe
because the three-dimensional homogeneity and isotropy of the 
bulk must be assumed {\it a priori}. In this paper we propose a 
mechanism by which a brane surrounded by AdS space 
arises naturally in such a way that the homogeneity and isotropy
of both the brane and the bulk are guaranteed. 
We postulate an initial false vacuum phase of $(4+1)$-dimensional
Minkowski or de Sitter space subsequently decaying 
to a true vacuum of anti-de Sitter space, assumed discretely 
degenerate. This decay takes place through bubble nucleation.
When two bubbles of the true AdS vacuum eventually collide,
because of the degeneracy of the true AdS vacuum, a brane (or domain wall) 
inevitably forms separating the two AdS phases. It is
on this brane that we live. The $SO(3,1)$ symmetry of the collision 
geometry ensures the three-dimensional 
spatial homogeneity and isotropy of the universe 
on the brane as well as of the bulk. In the semi-classical
$(\hbar \to 0)$ limit, this $SO(3,1)$ symmetry is exact. We sketch 
how the leading quantum corrections translate into cosmological
perturbations.

\end{abstract}

\section{Introduction}

We propose a cosmogony based on collisions of true 
anti-de Sitter (AdS) vacuum bubbles in $(4+1)$ dimensions expanding
at nearly the speed of light within a surrounding $(4+1)$-dimensional 
de Sitter (dS) or Minkowski (M) space false vacuum. The bubble collisions 
produce a braneworld universe very similar to the cosmogony 
with a $(3+1)$-dimensional, positive-tension brane surrounded by 
$(4+1)$-dimensional AdS space proposed by Randall and 
Sundrum\cite{randall,langloisa}.

Initially a $(4+1)$-dimensional spacetime consisting of 
either de Sitter space or Minkowski space is supposed. 
In the former case an initial epoch of $(4+1)$-dimensional 
`old' inflation\cite{guth} ensures a very nearly $SO(5,1)$ symmetric state 
prior to bubble nucleation, regardless of whatever 
departures from de Sitter space may have initially existed. 
The homogeneity and isotropy of the resulting 
$(3+1)$-dimensional braneworld universe is thus assured, 
as we shall explain in more detail. In the latter case, 
a metastable $(4+1)$-dimensional Minkowski state vacuum must be 
postulated at the outset; however, it is not at all implausible that
some as yet unknown theory of the initial conditions of the 
universe prefers empty Minkowski space. 

The false de Sitter or Minkowski vacuum decays through the nucleation 
by quantum tunnelling of bubbles filled with the lower energy true AdS vacuum
\cite{colemana,colemanb,russian}. The bubble wall separating the two phases
may take the form of either a brane or an accelerating domain wall.
We postulate that the AdS vacuum is discretely degenerate, so that the 
energy from the collision of two bubbles is not entirely transformed
into energy dispersed into the $(4+1)$ dimensions. 
In the case of a degenerate AdS vacuum, when the two colliding bubbles 
contain differing AdS phases, after the collision at least part of the energy
is transferred to a brane (or domain wall) that must mediate 
between the two phases. This is energy that remains localized
in the fifth dimension. In this paper we shall call this
brane (or domain wall) our {\it local brane} because this is where
the $(3+1)$-dimensional universe familiar to us resides.

To the extent that our universe has a violent beginning 
resulting from the collision of branes, the scenario presented
here has much in common with the brane inflation proposed by Dvali
and Tye\cite{dvali} and the ekpyrotic universe recently proposed 
by Khoury et al.\cite{turok}; however, the physics by which preferred
initial conditions are determined is quite different. 
The scenario proposed here also bears some similarities to the 
work of Gorsky and Selivanov\cite{gorsky}.
Perkins\cite{perkins} considered a braneworld scenario in which our universe
is situated on a bubble wall. However, in his scenario bubble
collisions are regarded as catastrophic. The dynamics of bubble collisions have
been studied by Guth and Weinberg\cite{weinberg},
Hawking, Moss and Stewart\cite{hawking},
and Wu\cite{chao}.

Before embarking on a detailed description of the colliding bubble 
scenario, we first highlight some of the problems arising from the 
assumption of a bulk with a negative cosmological constant. 
These difficulties, which render many braneworld cosmogonies problematic,
are avoided in the scenario proposed here because of the 
presence of a prior epoch of de Sitter or Minkowski space. 
Most braneworld models, including those with inflation on the brane, 
are plagued by the same horizon and smoothness problems present in 
non-inflationary cosmogonies but in $(4+1)$ rather than $(3+1)$ dimensions.  
The persistence of very near spatial homogeneity and isotropy on the brane 
requires that the bulk at the outset be so very nearly three-dimensionally homogeneous 
and isotropic\cite{trodden}. 
Otherwise, through gravity an inhomogeneous bulk inevitably 
induces inhomogeneities on the brane. A successful braneworld 
cosmology must therefore explain why the bulk was very nearly homogeneous and 
isotropic at the beginning.  A mechanism that merely smooths an initially 
inhomogeneous brane embedded in pristine AdS space, such as brane inflation, 
does not suffice because the necessary bulk homogeneity and isotropy 
must be put in by hand. 

Anti de Sitter space, or more broadly any spacetime with the stress-energy of 
a negative cosmological constant, lacks the ability to erase small initial 
perturbations from homogeneity and isotropy. For the case of a positive 
cosmological constant departures from homogeneity and isotropy rapidly disappear
as the universe expands. This is what provides the magic of inflation, by which 
perturbations are rapidly stretched to scales too large to be observed so that
after a rather modest amount of expansion one is left with essentially 
stretched vacuum. One could perversely attempt to postulate some sort of 
fractal state for which no amount of inflationary expansion would yield a 
homogeneous and isotropic state on the scales of interest, but such an 
initial state would entail an infinite energy density and thus can be excluded,
even under the most generous restrictions on admissible initial states. 

\begin{figure}
\begin{picture}(600,150)
\put(80,80){\leavevmode\epsfxsize=2in\epsfbox{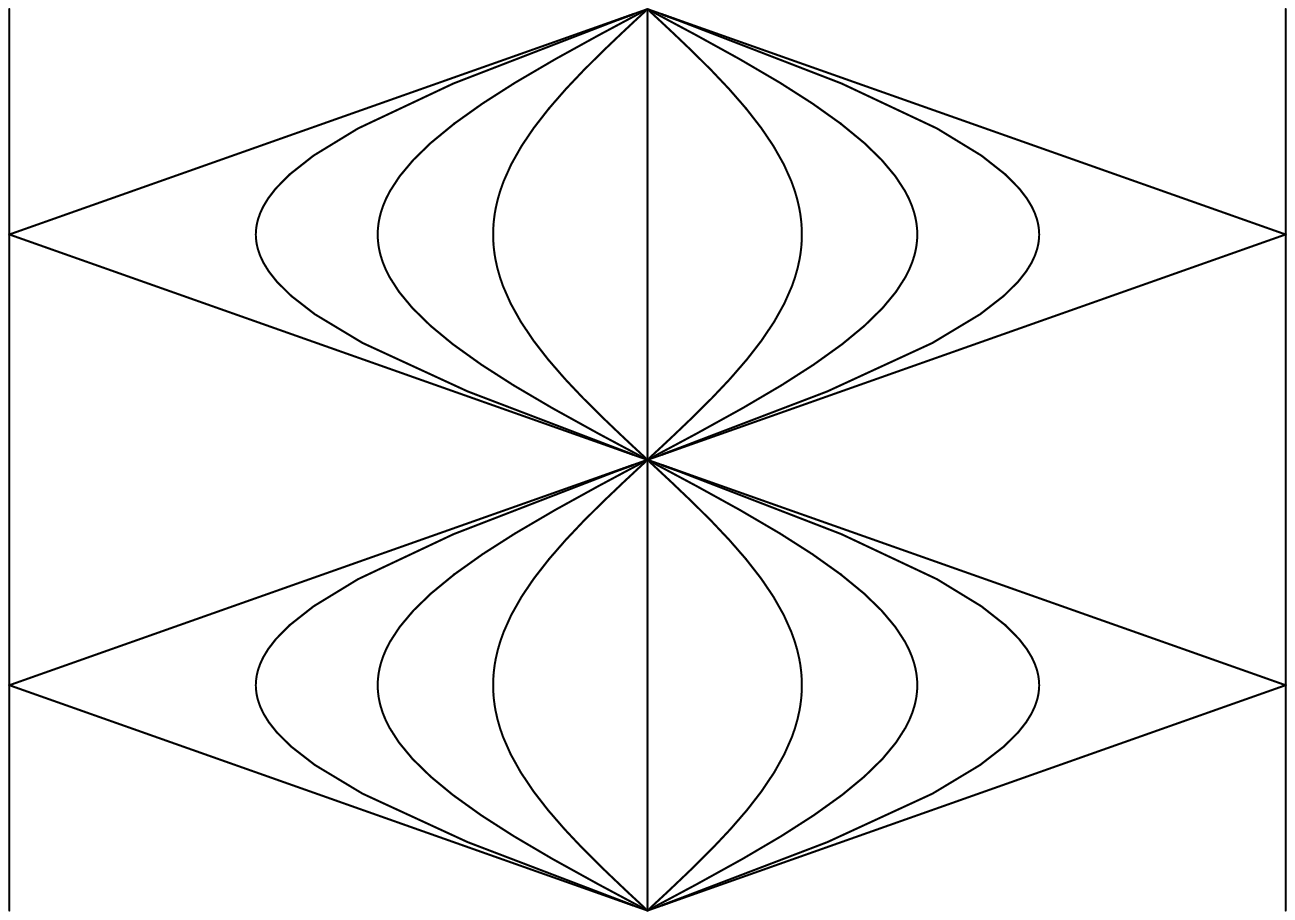}}
\put(150,70){P}
\put(300,65){\leavevmode\epsfxsize=2in\epsfbox{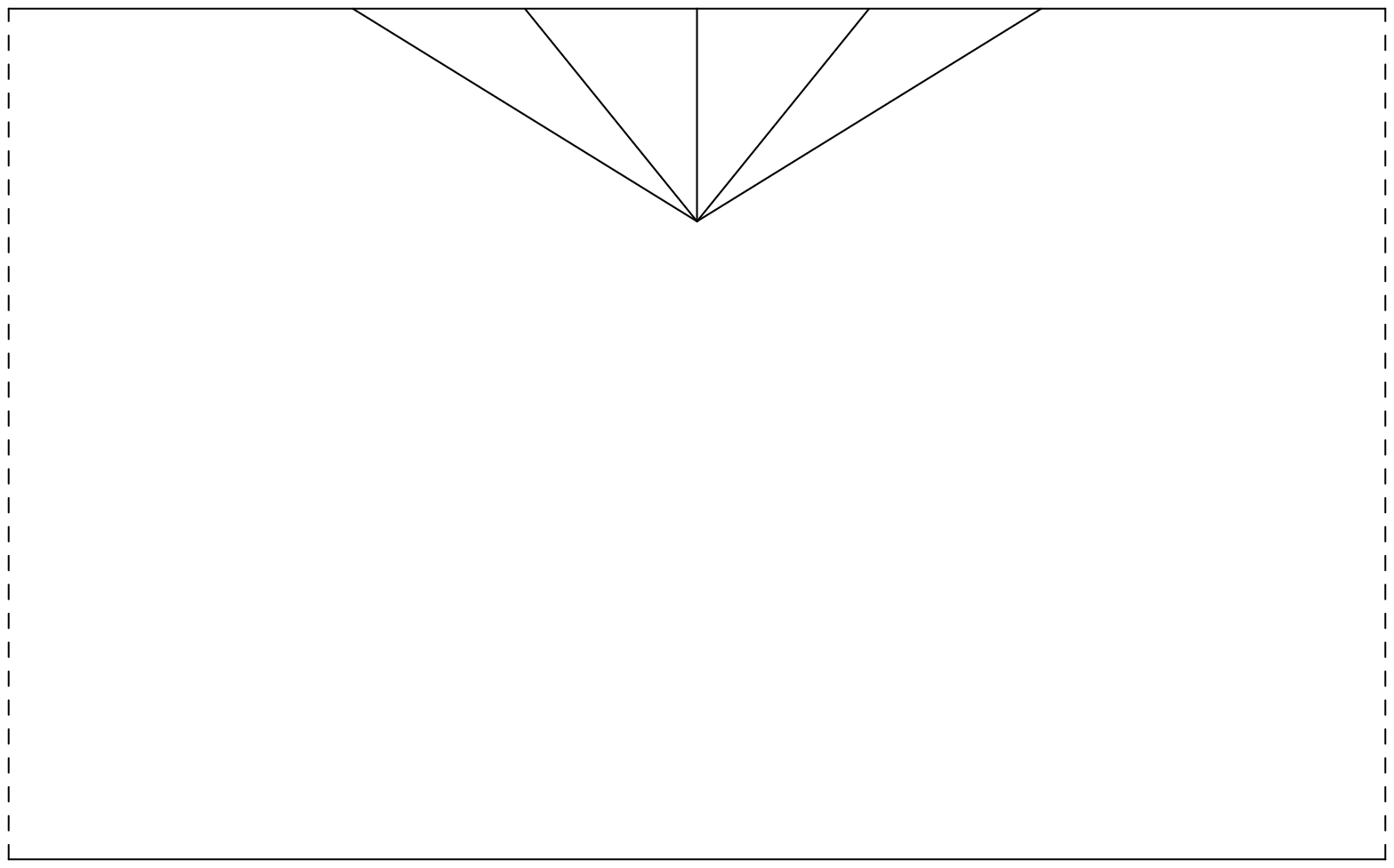}}
\put(368,135){P}
\put(150,40){(a)}
\put(368,40){(b)}
\end{picture}
\vskip -25pt
\caption{
{\bf Differing evolution of timelike geodesics in
anti de Sitter and de Sitter space.} The left panel (a) shows
a conformal diagram for anti-de Sitter space, which has the form
of an infinite vertical strip of finite thickness. The horizontal
and vertical directions indicate space and time, respectively,
and null geodesics travel obliquely at 45 degrees. The right
panel (b) shows the conformal diagram of de Sitter space which has the
form of a cylinder of finite height. The dashed vertical boundaries
are identified. In both panels the forward timelike geodesics of
a spacetime point $P$ are indicated, as well as the asymptotic light
cones forming the boundary of the causal future of $P.$ In anti-de Sitter
space the timelike geodesics periodically refocus {ad infinitum.}
By contrast, in de Sitter space the geodesics diverge, eventually
losing causal contact with each other.
}
\label{fig:1}
\end{figure}

The differing evolution of dS and AdS space is readily illustrated
by considering the family of timelike geodesics emanating from an 
arbitrary point $P$ 
in the spacetime, as indicated in Fig.~1. One might for example 
interpret these geodesics as the worldlines of the shrapnel from an 
exploding bomb! In both cases the trajectories initially diverge in proportion 
to their relative velocities, just as in a Milne universe 
(which is but another coordinatization of flat Minkowski space). 
However, after a proper time comparable to the curvature radius, the 
trajectories in AdS start to converge, eventually refocusing to a point 
(where the bomb momentarily re-assembles itself!). This sequence of
divergence and reconvergence repeats itself {\it ad infinitum.}
In de Sitter space, however, the nonvanishing spacetime curvature has precisely
the opposite effect. Rather than re-converging, the initial linear
divergence of the trajectories accelerates, so that eventually the 
pieces of shrapnel lose causal contact with each other. Exponentially 
inflating spacetime inserts itself between the fragments. In summary,
de Sitter space loses its ``hair", while anti de-Sitter space does not. 

In addition to the persistence of the irregularities in the manner  
just described, anti-de Sitter space is plagued with a bizarre
causal structure. As indicated in Fig. 1(a), maximally extended 
AdS space is bounded by timelike boundaries at infinity from which and to which
information flows. It does not make sense to postulate eternal
AdS without some theory of appropriate boundary conditions on these edges
or on the Cauchy horizons that result when one attempts to limit
consideration to a subspace of the maximally extended spacetime.
In the original Randall-Sundrum proposal (whose causal structure
is indicated in Fig. 2), the usual Randall-Sundrum coordinates 
\begin{equation}
ds^2=dy^2+\exp [2y] 
\cdot \Bigl[ -{dx_0}^2 + {dx_1}^2 +{dx_2}^2 +{dx_3}^2 \Bigr]
\label{boo}
\end{equation}
cover only a minute portion of maximally extended AdS space. 
The coordinate patch covered by (\ref{boo}) forms a 
globally hyperbolic subspace of maximally extended 
anti de Sitter space---that is,
initial data on a slice of constant cosmic time in the 
Randall-Sundrum coordinates is not constrained by any
consistency conditions and completely suffices to determine 
the fields in the triangular region covered by these coordinates.
But the lower light-like boundary constitutes a Cauchy horizon,
and one may legitimately inquire, what principle determines the initial
conditions on this boundary? And if they are trivial, as is often
assumed, why is this so? 
Although Fig.~2 illustrates the special case of
a static Randall-Sundrum universe, the lower Cauchy 
horizon persists in Randall-Sundrum cosmological models of an expanding
universe. 

\begin{figure}[h]
\vskip -15pt
\begin{picture}(600,150)
\put(200,20){\leavevmode\epsfxsize=2in\epsfbox{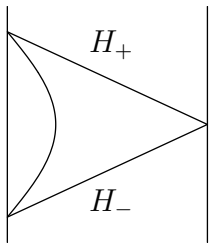}}
\put(255,48){$H_-$}
\put(255,108){$H_+$}
\end{picture}
\vskip -25pt
\caption{{\bf Causal structure of the single-brane Randall-Sundrum
braneworld spacetime.} The surfaces of constant time in the
Randall-Sundrum coordinates are generated by the family of all spacelike
geodesics emanating from a certain fixed point on the conformal boundary. The
worldlines of constant transverse coordinate (i.e., the ``fifth dimension" of
the Randall-Sundrum scenario) represent uniformly accelerating observers,
all with the same uniform acceleration away from this point. The Cauchy
horizons $H_-$ and $H_+$ coincide with the past and future boundaries of the
region covered by these coordinates.
}
\end{figure}

In the present proposal AdS bubbles arise
through the decay of a false de Sitter space or Minkowski
space vacuum. The AdS space that emanates inside
the bubble is produced in a precise and predictable way,
with quantum fluctuations that are predictable and calculable.
The problems described above are avoided. In the next section,
we describe the geometry and dynamics of the production and collisions of
AdS bubbles, explaining why in the semi-classical
$(\hbar \to 0)$ limit the resulting brane universe is 
homogeneous and isotropic. 
In section III we turn to the leading quantum corrections to this
picture, presenting a simplified calculation of the quantum 
fluctuations, which in our universe translate into a spectrum
of Gaussian linearized cosmological perturbations.
In the final section we present some concluding remarks. 

\section{AdS From Colliding Bubbles}

The possibility has been previously advanced that the true 
vacuum might not coincide with what we commonly perceive as  
the true vacuum. That is, rather than being either empty Minkowski 
space or de Sitter space with a remarkably small positive 
cosmological constant, the true vacuum might take the form of some lower energy 
state with a negative cosmological constant. If this were true, 
we would live in a metastable false vacuum state 
susceptible to decay to the true vacuum through bubble nucleation. 
Phenomenologically, given the observed persistence of our universe, 
an approximate upper bound on the rate $\Gamma $ at which bubbles of 
true AdS vacuum spontaneously nucleate can be established, but it 
is not possible to reject this possibility altogether.  

\begin{figure}[h]
%\vskip -15pt
\begin{picture}(600,150)
\put(125,20){\leavevmode\epsfxsize=3in\epsfbox{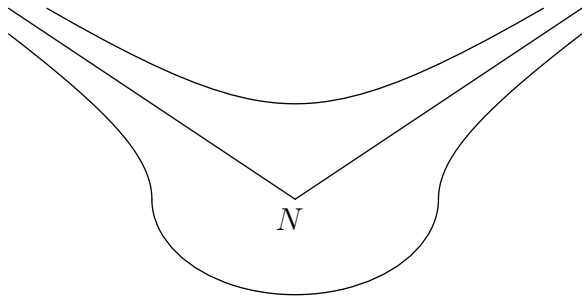}}
\put(226,63){$N$}
\end{picture}
\vskip -25pt

\caption{{\bf Dynamics of vacuum decay through the nucleation of a
single bubble.} Above is shown the nucleation through quantum tunnelling and
the subsequent classical evolution of a single bubble (for simplicity in
Minkowski space).  As before, time increases in the vertical direction
and the horizontal direction indicates one of the four spatial directions.
The scalar field is constant on the solid curves. In the lower part of the diagram
the nucleation of the bubble is represented by the concentric circles.
However, since the inherently quantum mechanical process
that produces the initial critical bubble cannot be observed without
altering its outcome, it is best to regard this part of the evolution
as a quantum mechanical black box whose inner workings must remain hidden.
During the subsequent classical expansion of the bubble, the velocity of the 
bubble wall approaches $c.$}
\end{figure}

A manifestly covariant description of the dynamics of false vacuum was 
given by Sidney Coleman, first ignoring gravity \cite{colemana}
and then extended to include the gravitational corrections in work
with F.~de Luccia \cite{colemanb}. Important prior work
is contained in \cite{russian}. This process is illustrated
in Fig.~3. We summarize below the principal
results of these papers to the extent that they are needed here
and refer the reader to the original papers
for a more detailed and rigorous discussion.

False vacuum decay takes place at zero temperature, or said
another way, from an initial state no preferred time 
direction. The consequences of the lack of a
preferred time direction are profound. They render false
vacuum decay qualitatively different from the more familiar
thermal tunnelling, which enjoys considerably less
symmetry due to the fact that a thermal state singles out a 
preferred time direction. For false vacuum decay in
$(d+1)$ dimensions, the resulting classical expanding bubble
solution possesses an $SO(d,1)$ symmetry. The symmetry group in the 
absence of any bubbles
[which in the case of de Sitter space would be $SO(d+1,1)$]
is broken by the presence of a single bubble
to $SO(d,1)$, the subgroup of transformations
that leaves invariant a spacetime point known as the {\it nucleation center.}

For a spacetime with two bubbles, the resulting symmetry is 
further reduced, but considerable residual symmetry
remains. Suppose that two bubbles nucleate at spacelike separated 
nucleation centers $N_L$ and $N_R,$ where $L$ and $R$ denote left
and right. This separation must be 
spacelike, for otherwise one bubble would nucleate within the other.
For two bubbles the solution remains symmetric under the subgroup of 
those transformations that leave invariant 
the line (or spacelike geodesic) passing through $N_L$ and $N_R.$ For
a pair of colliding bubbles nucleating in (4+1)-dimensional dS space, 
the $SO(5,1)$ symmetry breaks to $SO(3,1).$ This residual
symmetry has the following consequences. First, one may always choose
a coordinate system in which the two bubbles nucleate 
at the same time. Hence, unlike for thermal tunnelling, here
it is not meaningful to ask which of the two bubbles is the bigger one.
Moreover, once a coordinate choice is made in which
the bubbles nucleate simultaneously, substantial 
residual symmetry remains. While in a particular coordinate
system the bubbles first collide at a given spacetime point $P,$
for any other point $P'$ of the locus of points where
the bubbles collide, a coordinate transformation exists 
such that the bubbles first collide at $P'.$ It is this symmetry
mapping $P$ into $P'$ that is responsible for the 
three-dimensional spatial homogeneity 
and isotropy of the universe on the local brane.

\begin{figure}[t]
\begin{picture}(600,150)
\put(80,30){\leavevmode\epsfxsize=2in\epsfbox{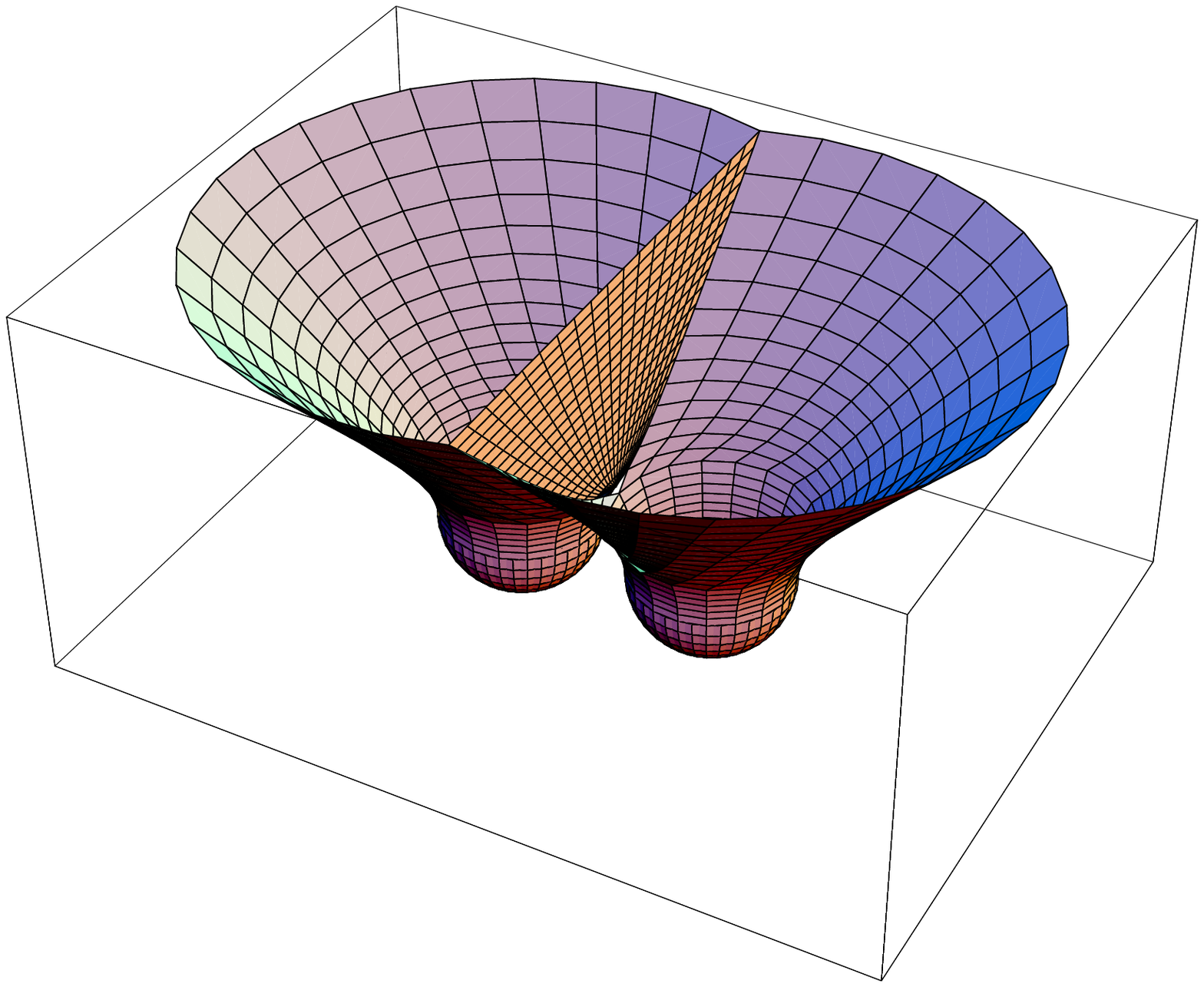}}
\put(300,30){\leavevmode\epsfxsize=2in\epsfbox{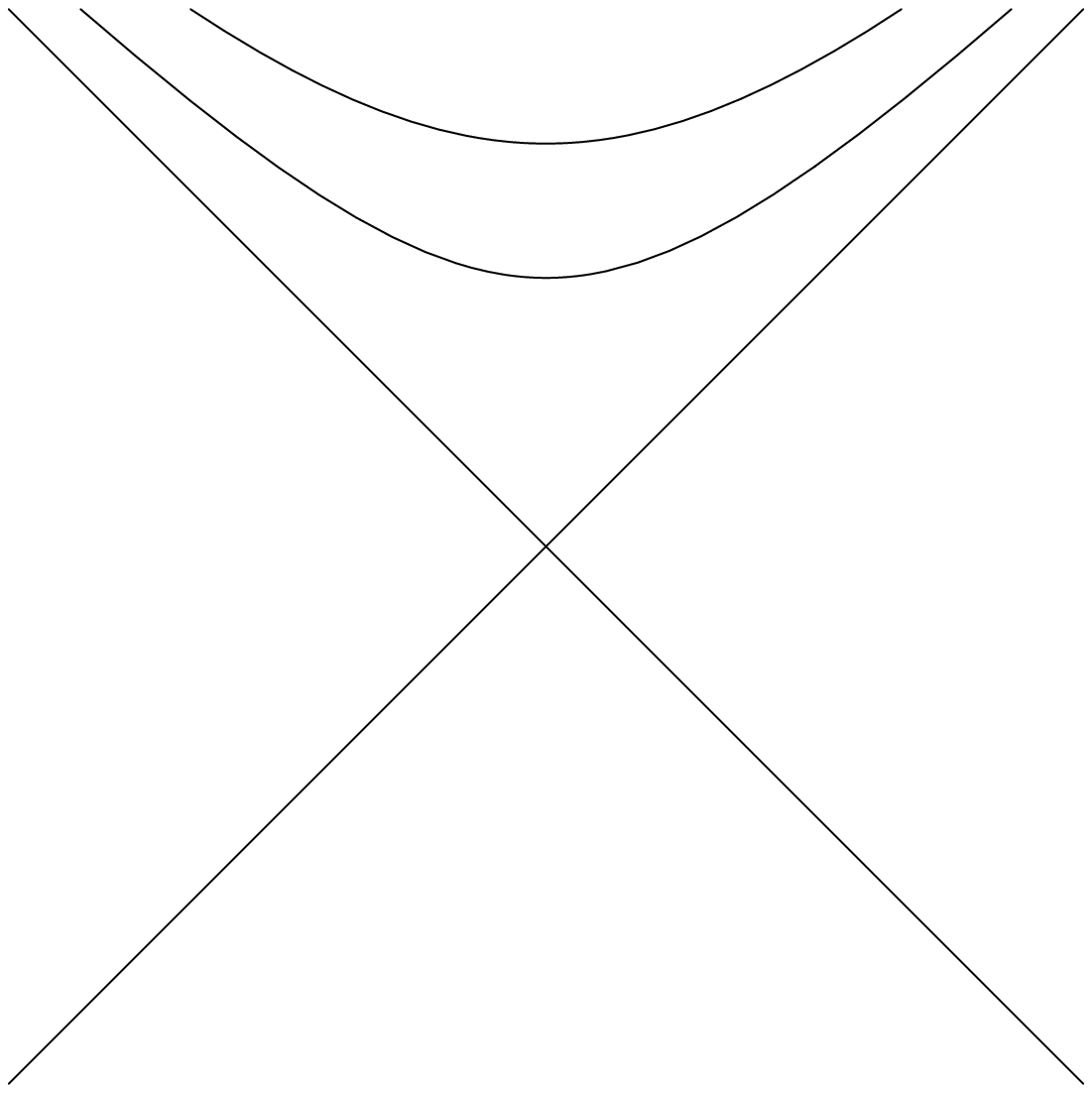}}
\put(365,107){$C$}
\put(380,87){$M$}
\end{picture}
\caption{{\bf The geometry of the collision of two bubbles.}
The left panel indicates the collision of two bubbles, represented
in the thin wall limit with (2+1) dimensions shown.
The vertical direction represents time.
The right panel indicates a cross section of the
plane exactly midway between the two nucleation centers.}
\end{figure}

We now turn to a more detailed consideration of what happens during
the bubble collision.  For vacuum decay with a single scalar 
field where the AdS vacuum is nondegenerate, the energy 
of the colliding bubble walls, absent some good reason to the contrary, 
dissipates in the fifth dimension (the direction parallel to the line connecting
the two nucleation centers) but in an $SO(3,1)$ symmetric way, 
much as in the initial stages of thermalization first envisaged for 
`old'\cite{guth} or `extended'\cite{steinhardt} inflation. However, if the AdS vacuum
is finitely degenerate (in the simplest case with two such AdS 
vacua related by a $Z_2$ symmetry), topology demands that a domain wall form
after the bubble collision to separate the two distinct AdS domains when 
the colliding bubbles contain differing AdS phases. 
While energy that disperses in the 
fifth dimension could as well be produced in the collision, 
topology requires that a domain wall form to mediate between the two
AdS states. This wall, which we call our {\it local brane} (on which we
live) is at rest in the center of mass frame of the colliding 
bubble. Of the kinetic energy left over after this domain wall has been
formed, a part is expected to stick to the brane (and to be confined to it, as is 
typically assumed in the Randall-Sundrum scenario) and another part is expected to 
disperse into the bulk. The energy dispersed into the bulk, however,
is $SO(3,1)$ symmetric, and therefore does not induce any irregularities on
the brane. Moreover, this energy does not fall back onto the brane, because 
when the gravitation of the brane is taken into account, the brane accelerates
away from this symmetric, dispersive debris. 

Fig.~4 shows schematically first a (2+1)-dimensional representation
of the colliding bubble geometry in the left panel and then in the 
right panel a cut-away of the surface of equal proper distance from
the two nucleation centers. The point $M$ is the midpoint of $N_L$ and $N_R.$
The curve labeled $C$ indicates the 
line along which the two bubbles collide. In the section on the 
right, several hyperbolic coordinate patches are generated by 
the $SO(3,1)$ symmetry separated by the backward and forward lightcones on $M.$ 
Points along the solid curves are rendered equivalent 
by this symmetry. These are lines of constant cosmic temperature
on our local brane, which cools as the universe expands. In
the full (3+1)-dimensional case, these curves are three-dimensional
spacelike hyperboloids of constant negative spatial curvature. 

\begin{figure}[h]
%\vskip -15pt
\begin{picture}(600,150)
\put(125,20){\leavevmode\epsfxsize=2.5in\epsfbox{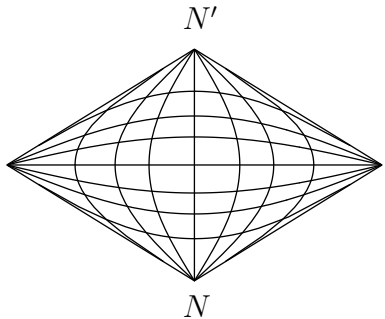}}
\put(211,38){$N$}
\put(211,147){$N'$}
\end{picture}
\vskip -25pt
\caption{{\bf Fate of a single AdS bubble.} The bubble interior
with the geometry of AdS space is indicated. The scalar field is
constant along the surfaces indicated by the solid curves. $N$
is the nucleation center and $N'$ is the point at which the
timelike geodesics emanating from $N$ first reconverge. The surfaces
on which the scalar field is constant are normal to these geodesics.}
\end{figure}

It has been suggested by Coleman that isolated AdS bubbles generically
collapse into black holes because of the $SO(4,1)$ symmetry of
the perfect classical expanding bubble solution. The argument, 
which is closely related to the perfect refocusing of timelike
geodesics emanating from a point described above, is as 
follows.  The universe inside an AdS bubble is a hyperbolic
universe that recollapses after a finite amount of time. If the background 
stress-energy inside the bubble were that of a perfect negative cosmological
constant, this would pose no problem. The resulting `Big Crunch' 
would be nothing but a coordinate artifact, as indicated in Fig. 5.
However, if the scalar field undergoing tunnelling has not reached
the true vacuum by the light cone $L$ (which it never does),
a singularity in the evolution of the scalar field on the light cone $L'$ 
generally results.  In both cases the behavior of the scalar field near 
the lightcones is described by a second-order, Bessel-like ordinary
differential equation having one regular and one singular solution. 
On $L$ it is clearly correct to choose the regular solution. This is
the initial condition that results from the Euclidean instanton. 
But unless the potential is extremely finely tuned, upon propagating to 
$L',$ at least a small admixture of the divergent, irregular solution 
will be present, causing the scalar field kinetic energy to diverge. In the 
case of colliding bubbles, however, the underlying symmetry that led to 
the divergence is broken because the collision generically sends out a wave 
that spoils the finely tuned convergence of the scalar field that
led to black hole formation. Thus the AdS space inside the bubbles is allowed 
to persist. 

\begin{figure}[h]
\begin{picture}(600,150)
\put(80,30){\leavevmode\epsfxsize=2in\epsfbox{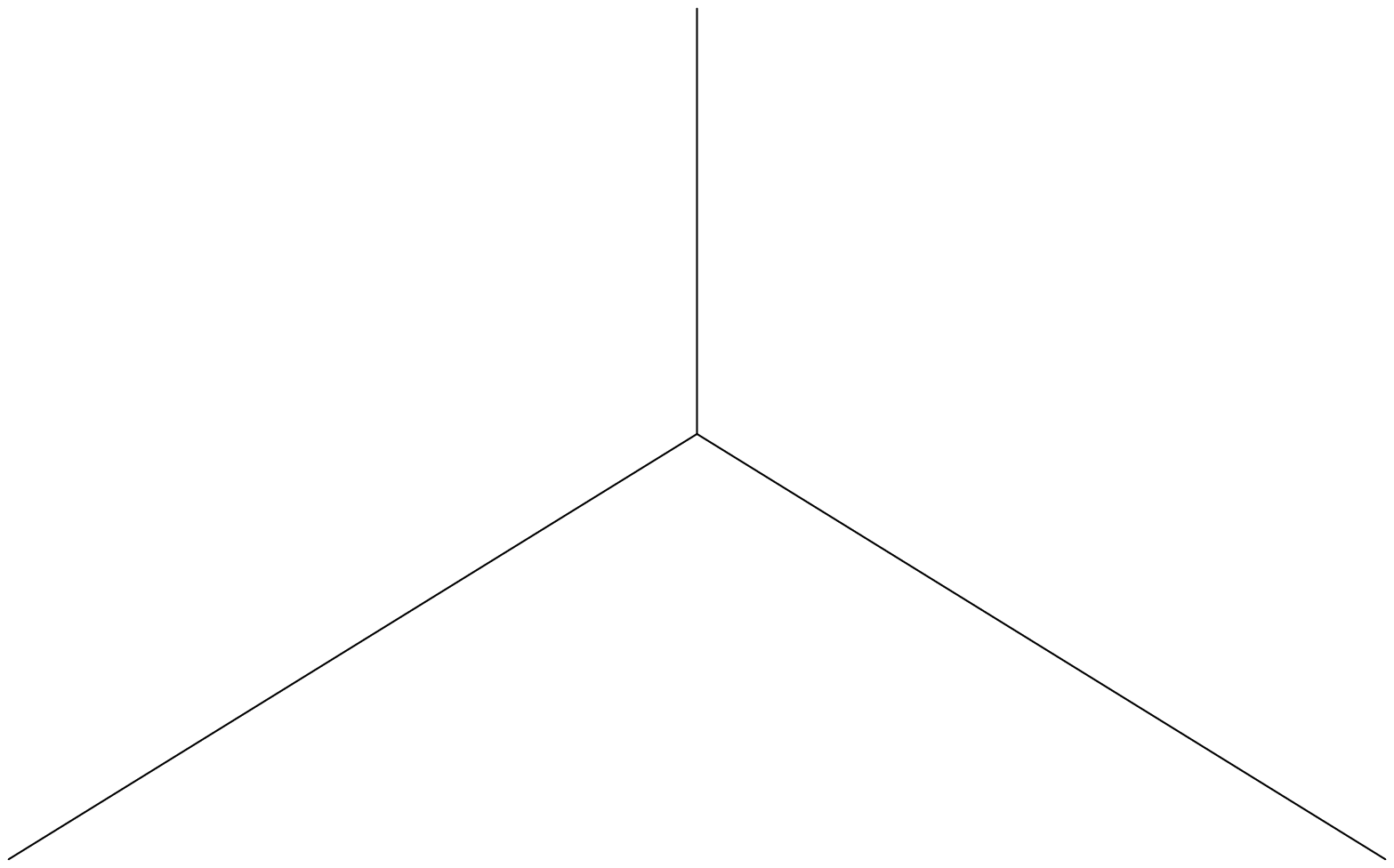}}
\put(79,72){$\rho _L\bar {\bf u}_L$}
\put(95,60){\vector(3,2){15}}
\put(200,72){$\rho _R\bar {\bf u}_R$}
\put(160,105){$\rho _F\bar {\bf u}_F$}
\put(210,60){\vector(-3,2){15}}
\put(145,100){\vector(0,2){15}}
\put(300,30){\leavevmode\epsfxsize=2in\epsfbox{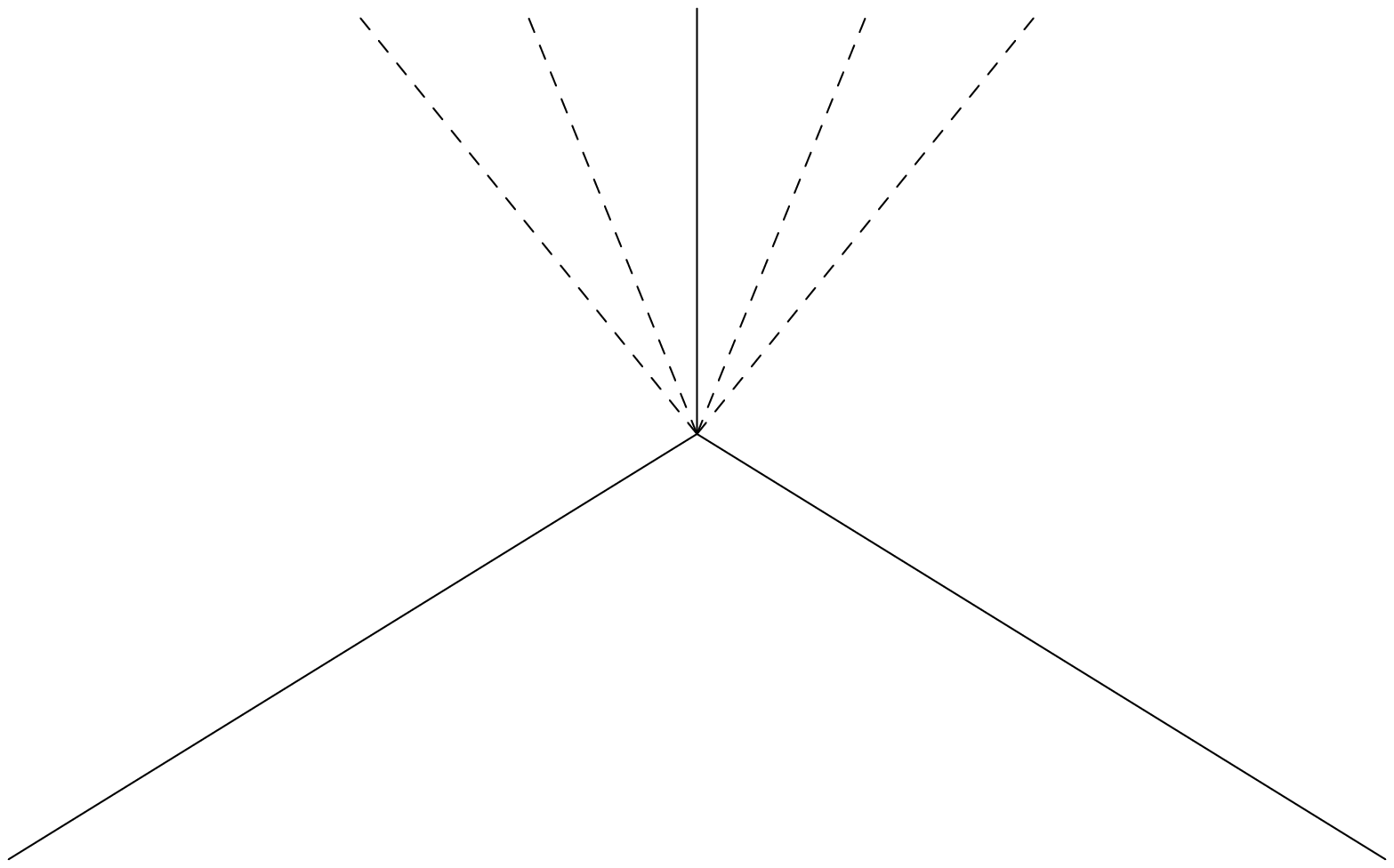}}
\end{picture}
\vskip -25pt
\caption{{\bf Stress-energy conservation during brane collisions.}
Collisions or decays of branes may be represented using
a sort of Feynman diagram in time and the transverse spatial
dimension. The three homogeneous spatial directions are suppressed.
The vectors $\rho \bar {\bf u},$ where $\rho $ is the density on
the brane in the brane rest frame and $\bar {\bf u}$ is the vector
tangent to the brane, must all sum to zero at the vertex. In
the left panel, the collision of two branes, with all the available energy
is deposited onto a single brane in the final state, is represented.
In the right panel, the case where some of this energy is emitted
as dispersive waves (shown by the dashed trajectories) is indicated.}
\end{figure}

To simplify the analysis, we idealize the bubble walls as infinitely
thin and assume that upon colliding the bubbles transfer all their
available energy onto an infinitely thin brane, with all excess 
energy converted into radiation and matter confined to the brane. 
The collision geometry is indicated in Fig. 6. The subsequent evolution
of the brane depend on the equation of state on our brane, which
we take to be arbitrary, since the considerations presented in
this paper do not depend on its details.

Since bubbles nucleate stochastically, at a rate $\Gamma $
with the dimension of inverse volume inverse time, the proper distance between 
nucleation centers is a random variable. Consequently,
the spatial curvature of the resulting intermediate brane
universe varies between bubble pairs. In this scenario it is 
essential that bubble collisions are rare. A collision with a third 
bubble would be catastrophic; a wave of energy would move toward us 
at very nearly the speed of light striking us with essentially no warning.   
That this has not yet happened is a most trivial application of the anthropic 
principle. In the case of bubbles expanding in Minkowski space $(M^5),$ if
the nucleation rate $\Gamma $ does not vary with time, the bubbles will
all eventually percolate. Therefore 
the exterior $M^5$ space could not have persisted 
infinitely far into the past unless some mechanism, such as a variant
of that of extended inflation\cite{steinhardt}, is postulated to render $M^5$ 
eternal into the past by making $\Gamma $ vanish in limit of 
the infinite past.  This percolation, however, does not occur for bubbles 
expanding in $dS^5$ for the small nucleation rates 
of interest here.\cite{weinberg}

\section{ Quantum Corrections: Generation of Gaussian Cosmological
Perturbations}

In the previous section we demonstrated how a homogeneous and isotropic
universe can arise from the collision of two expanding AdS bubbles.
We employed the semi-classical $(\hbar \to 0)$ limit in which prior to colliding
each bubble possesses an exact $SO(4,1)$ symmetry about its nucleation center,
because in the semi-classical limit 
fluctuations about the configuration of least Euclidean action describing 
the bubble nucleation process are suppressed as well as the quantum fluctuations 
of the wall and of the surrounding fields afterward. In this limit one obtains
an absolutely homogeneous and isotropic universe, quite unlike the
one that we observe. Quantum corrections, however, alter this picture. 
The leading order corrections in $\hbar $ yield a calculable spectrum
of linearized Gaussian fluctuations. These are the usual Gaussian cosmological 
perturbations. 

For calculating the cosmological perturbations, the Bunch-Davies vacuum
of de Sitter space (or the Minkowski space vacuum for the case of bubbles 
nucleating in Minkowski space) define a natural set of initial conditions. The 
Bunch-Davies vacuum is an attractor, so an initial state deviating from
this state evolves to become successively better approximated 
by the Bunch-Davies vacuum.  A full calculation of the perturbations is
postponed until a later paper\cite{bucher}. Here we limit ourselves to a simplified
qualitative description ignoring gravitational back reaction and assuming 
infinitely thin bubbles to illustrate the underlying
physical processes. 

The quantum state for the fluctuations of a thin wall bubble about the
perfect $SO(4,1)$ expanding bubble solution for a bubble arising from 
false vacuum decay was first elucidated by J. Garriga and 
A. Vilenkin\cite{garrigaa}.  In the thin wall approximation, with the gravitational 
back reaction of the perturbations ignored, the only available degree of 
freedom consists of normal displacements of the bubble wall, which may be 
described as a scalar field localized on the bubble wall itself. 
We consider the perfect $SO(4,1)$ 
symmetric expanding bubble (which has the geometry of de Sitter space). 
Displacements along the outward normal are described by a free scalar
field of mass $m^2=-4H^2.$ The quantum state of this field must obey the 
same $SO(4,1)$ symmetry as the classical expanding bubble solution. 
One might at first sight admit the possibility that bubble nucleation could somehow
spontaneously single out a preferred time direction. That this is
not possible can be demonstrated by contradiction. 
Suppose that such a choice of preferred time direction were in fact 
made. Then all such choices must be equally weighted, 
according to a Lorentz invariant measure. The calculation of
the vacuum decay rate would contain a factor consisting of an integration 
over the infinite hyperbolic domain (with the geometry of $H^3$)
of all such possible choices, thus
implying an infinite false vacuum decay rate, a conclusion 
which is clearly absurd.  The $SO(4,1)$ invariance
of the quantum state of the fluctuations suffices to completely fix
this state. It is described by the Bunch-Davies vacuum of the de Sitter
space of the expanding bubble wall.

\begin{figure}[h]
\begin{picture}(600,150)
\put(80,30){\leavevmode\epsfxsize=2in\epsfbox{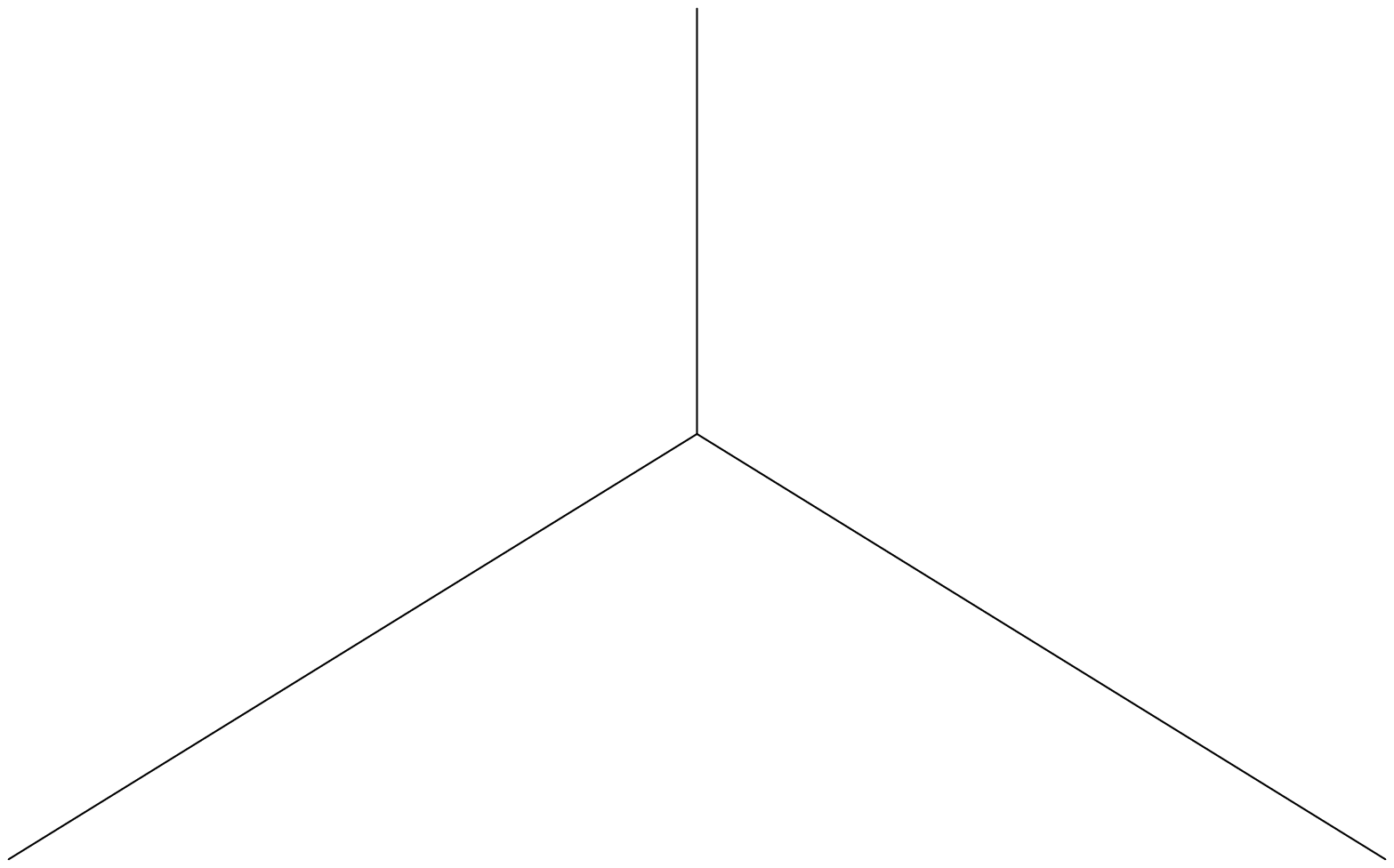}}
\put(120,63){\vector(1,-1){15}}
\put(111,49){$\chi _L$}
\put(180,63){\vector(-1,-1){15}}
\put(180,49){$\chi _R$}
\put(165,130){\vector(1,0){15}}
\put(182,129){$\chi _-$}
\put(160,125){\vector(0,-1){15}}
\put(157,100){$\chi _+$}
\put(300,-20){\leavevmode\epsfxsize=2in\epsfbox{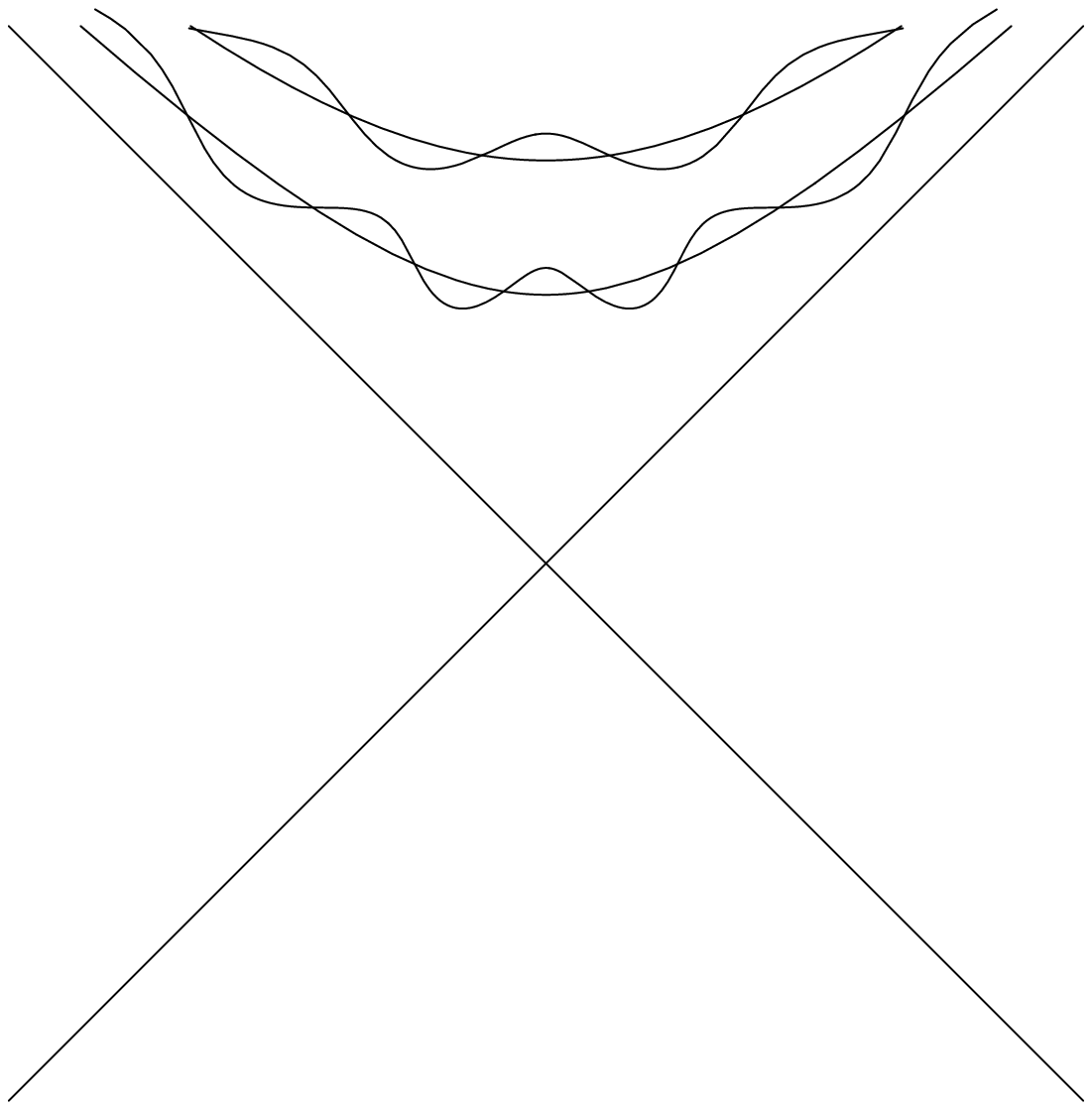}}
\end{picture}
\vskip -25pt
\caption{{\bf Perturbations in the thin brane approximation.}
The displacements $\chi _L$ and $\chi _R$ of the two expanding
bubble walls resolve into the displacements $\chi _+$ and
$\chi _-$ of the local brane. The mode $\chi _-$ represents
lateral distortion of the local brane. The mode $\chi _+$
advances or retards the moment of bubble collision.
The right panel illustrates the distortion of the surfaces of
constant cosmic temperature of the local brane.}
\end{figure}

Let $\chi _L$ and $\chi _R$ be the scalar fields just described
for the two colliding bubbles, using the sign convention that 
$\chi $ is positive for outward displacements. To analyse
how these displacements translate into perturbations of 
the brane that arises from the bubble collision, it is 
convenient to consider the linear combinations
\begin{eqnarray}
\chi _+&=&(\chi _L+\chi _R)/\sqrt{2},\nonumber\\ 
\chi _-&=&(\chi _L-\chi _R)/\sqrt{2}
\end{eqnarray}
at the instant of collision.  The mode $\chi _+$ 
temporally advances (or retards) the surface on which the bubbles 
collide leading to under and 
overdensities. The hyperboloids of constant cosmic temperature
are thus warped.
This mode translates into scalar density perturbations of the 
cosmology on the local brane.
The mode $\chi _-,$ on the other hand, displaces the surface
of collision in the normal direction---that is, spatially 
toward the one or the other bubble.

Although the geometry of the background solution is $Z_2$ 
symmetric, as in the Randall-Sundrum scenario, the $Z_2$ symmetry 
here is qualitatively different from the orbifold $Z_2$ symmetry 
postulated in the Randall-Sundrum proposal. In our proposal, 
both $Z_2$ even and $Z_2$ odd perturbations are allowed
because the degrees of freedom on one side of the brane
do not coincide with those on the other side. In
the Randall-Sundrum scenario with a single brane 
there is no bending mode because the relevant degrees of freedom 
have been decreed not to exist through the orbifold construction. 
In our case, this mode does in fact exist. The extrinsic curvature 
(relative to the outward normal) on the two sides need not coincide
because twice as many degrees of freedom are present. 

\section{Concluding Remarks}

We have demonstrated how the collision of two bubbles filled
with AdS space expanding in de Sitter space or Minkowski space
can give birth to a braneworld cosmology surrounded by
infinite anti de Sitter space, very similar to the single-brane 
Randall-Sundrun model.  In this colliding bubble scenario 
well-defined initial conditions naturally arise.
The smoothness and horizon problems in $(4+1)$ dimensions
are absent in this scenario. Although the considerations presented in this
paper apply equally well regardless of the equation of state
on the local brane produced after the bubble collision, the 
fact that inflation on the resulting (3+1) dimensional spatially
hyperbolic universe can altogether be avoided is intriguing. 
If sufficient energy is deposited on this brane after sufficient 
expansion of the initially nucleated bubble, 
$\Omega $ today can be very close to one.

We now consider some orders of magnitude. In the Randall-Sundrum
scenario (just as in compact five-dimensional Kaluza-Klein models),
an effective four-dimensional Planck mass $m_4$ large compared to the
five-dimensional Planck mass $m_5$ may be obtained by making the
size of the extra dimension $\ell $ large.  
Here we set $\hbar =c=1.$ In the Randall-Sundrum case
$\ell $ is the curvature radius of the AdS bulk. Since
${m_4}^2={m_5}^3\ell ,$  $m_4=m_5(m_5\ell )^{1/2}.$

The five-dimensional Einstein equation and Israel matching condition give
$\Lambda ={m_5}^3\ell ^{-2}$ and $\sigma ={m_5}^3\ell ^{-1},$ 
respectively, where $\Lambda $ is the five-dimensional negative cosmological
constant in the bulk and $\sigma $ is the four-dimensional cosmological
constant that would be required on the brane for it to have the geometry of 
four-dimensional Minkowski space $(M^4).$
The tension of the wall separating the AdS from
the dS phase and that of the local brane in general differ, but for
the order-of-magnitude analysis here we take them to 
coincide. It follows that the approximate size of the critical bubble
is $r\approx \sigma /\Lambda \approx \ell .$ The vacuum decay rate is
approximately $\Gamma =\ell ^{-5}\exp [-S_E]$ where
$S_E\approx \sigma r^4\approx (m_5\ell )^3=(m_4\ell )^2.$
An extra dimension large compared to the Planck 
scales makes the dimensionless Euclidean action large,
leading to an exponentially small bubble nucleation rate.
Therefore, a very substantial amount of expansion takes
place before bubble pairs collide, and three bubble collisions
are rare. The perturbations are of order $1/\sqrt{S_E}.$ 

We now consider the spatial curvature of the universe
on the local brane. The energy density on the brane 
produced at the bubble 
collision is approximately $E_C=(R/r)\sigma $ where
$\sigma \approx {m_5}^4(m_5\ell )^{-1}$ and 
$R$ is the distance between the nucleation centers of the 
two bubbles. As long as $(R/r)\lesssim (\ell /\ell _5)$
where ${m_5}^{-1}$ is the five-dimensional Planck length,
this energy density is sub-Planckian from the five-dimensional
point of view.\footnote{The case of trans-Planckian energy densities 
immediately after the brane collision, however, need not necessarily be discarded,
because the analysis of what happens afterward depends little on the details of 
how the universe cools after the collision. One might regard a brief trans-Planckian 
epoch after the collision as a sort of black box, much as re-heating at the end
of inflation is commonly regarded. One is able to compute the perturbations
from inflation with confidence despite our almost total 
ignorance of how re-heating occurs.}
At the collision $(1-\Omega _c)\approx (\ell /R)^3,$ which is 
exponentially small. 
The bubble pair separation $R$ is a random variable differing
from pair to pair. The average bubble pair separation $\bar R,$ however, 
may be estimated by setting $\Gamma \bar R^5\approx 1$ assuming bubbles expanding in
$M^5.$
A more detailed discussion of the probability distribution will appear
elsewhere.\cite{carvalho}
The factor $e^{-S_E}$ provides
a natural mechanism to adjust $\Omega _c $ so close to one that
$\Omega $ today remains very close to one without resort to
unnatural fine tuning. 

\vskip 12pt
\noindent
{\bf Acknowledgements:}
The author would like to thank John Barrow, Carla Carvalho, 
Gary Gibbons, David Langlois, Takahiro Tanaka, Neil Turok, 
and especially Jaume Garriga  for useful discussions.
MB was supported by Dennis Avery. The author would also like 
to thank the IAP in Paris and IFAE in Barcelona for their 
hospitality and the British Council in Madrid for a travel grant.

\end{document}